\documentclass{article}

\begin{document}

{\bf STRING FIELD THEORY FROM QUANTUM GRAVITY}

\bigskip

by: Louis Crane, Mathematics Department KSU

\bigskip

{\bf ABSTRACT:} Recent work on neutrino oscillations suggests that the
three generations of fermions in the standard model are related by
representations of the finite group A(4), the group of symmetries of
the tetrahedron. Motivated by this, we explore models which extend the
EPRL model for quantum gravity by coupling it to a bosonic quantum field of
representations of A(4). This coupling is possible because the representation
category of A(4) is a module category over the representation
categories used to construct the EPRL model. The vertex operators
which interchange vacua  in the resulting quantum field theory
reproduce the bosons and fermions of the standard model, up to issues
of symmetry breaking which we do not resolve. We are led to the
hypothesis that physical particles in nature represent vacuum changing
operators on a sea of invisible excitations which are only observable
in the A(4) representation labels which govern the horizontal symmetry
revealed in neutrino oscillations. The quantum field theory of the
A(4) representations is just the dual model on the extended lattice of
the Lie group $E_6$, as explained by the quantum Mckay correspondence
of Frenkel, Jing and Wang.

The coupled model can be thought of as string field theory, but
propagating on a discretized quantum spacetime rather than a classical
manifold. 

\bigskip

{\bf 0. INTRODUCTION}

\bigskip

In the last few years, a new development \cite{EPR} \cite{L} \cite
{EP} \cite{B} 
has largely resolved the
problems of the old BC model \cite{BC} for quantum gravity. It is now a natural
task to study extensions  of the EPRL model which would include
realistic matter fields. It would be extremely desirable to find
an algebraic extension of the EPRL model which was essentially unique
or at least had a small number of possibilities and which gave us the
standard model, rather than some random collection of particles and
fields.

Durring the same time frame, a series of delicate experiments
\cite{neutosc}
has
given us a detailed picture of the oscillations of the neutrinos of
the three generations (electron muon and tau neutrinos) into one
another. 
The oscillations turned
out to be much larger than the ones for quarks, and seem to be well
approximated by a form called the tribimaximal matrix \cite {TBM}.

The neutrino masses indicate that the neutrino has a right handed
component, called the sterile neutrino because it does not interact
under any force. This changes the form of the three generations of
fermions; each now has 16 Weyl fermions.

The new neutrino Physics is crucial to the search for a unified
theory; it is as if we have been working a jigsaw puzzle with a
missing piece. If the proposal in this paper is correct, the missing
piece was the most important one.

In explaining the tribimaximal matrix, a number of researchers have 
proposed that the
particles appear in representations of the discrete group A(4)
\cite{Ma}, 
the
alternating group on 4 letters, and that the interactions with the
Higgs particle be invariant under this symmetry. This reproduces the
tribimaximal form very nicely.

The fact that A(4) has three one dimensional representations and one
irreducible three dimensional representation allows the quark and
lepton mixing matrices to be very different. Quarks and neutrinos get
different A(4) representational labels in this approach.

The A(4) symmetry cannot be explained as a broken or residual symmetry
from  a Lie group \cite{DSP}, it can only be understood as a
fundamental 
symmetry
of nature. This lends itself much more naturally to coupling to a
discrete model such as EPRL rather than to a continuum Yang Mills
theory.

The EPRL model, like the BC model which preceeded it, is
constructed from the representation categories of Lie groups, which
are used to assign quantum geometrical variables to the faces of a
four dimensional simplicial complex. The tensor structure of the
representation categories is used in a natural way to construct the
model. The EPRL model, in particular, uses the representation
categories of SO(3) and SO(3,1), together with a functor connecting
them \cite{Me}.
 
The group A(4) which was interesting to the neutrino physicists is
also the group of symmetries of the tetrahedron. The author Ma
recognized this fact in the title of one of his papers on the
subject ''Plato's Fire'' \cite{PF}. For this reason, its representation category
has a very special relationship with the two categories in EPRL: they
have a tensor action on it. This is expressed mathematically by saying
that it is a module category \cite{Eting}.

The reason the representation category of A(4) is a module category
over the representation category of SO(3) is that A(4) is a subgroup
of SO(3). The discrete subgroups of SO(3) form a very short list,
there are only 3 interesting ones A(4), S(4), and A(5), corresponding
to the symmetry groups of the tetrahedron, cube and icosahedron.  

Thus, the group which appears as a symmetry in neutrino oscillations
is one of a very small number whose representations can be coupled to the tensor
structure of the EPRL model.

Now since SO(3,1) is a noncompact group, it cannot act unitarily on
the Hilbert space generated by the irreducible representations of
A(4).

This means that we can not create a unitary state sum model by
coupling a single representation of A(4) to each tetrahedron in the
triangulated spacetime of the EPRL model. Instead, we must have a
complete copy of a Fock space $F_t$ for Rep(A(4)). The work of Frenkel
Jing and Wang \cite {FJW} gives us a natural construction for $F_t$, 
which we discuss
below. 

The fact that we must have a complete Fock space at each site is
analogous to the Klein paradox in ordinary quantum field theory. The
EPRL model requires us to act on states by an arbitrary element of
SO(3,1), therefore by a boost. We must allow for pair production in
our case as well.

Now the space $F_t$ has extremely interesting and powerful structure. FJW
showed that it is isomorphic to the Fock space of the dual model on
the extended root lattice of the Lie group $E_6$. (The appearance of
this Lie group is an example of the Mckay correspondence \cite{Mckay}).

This means it has the following structures:
\bigskip

{\bf STRUCTURE OF THE FOCK SPACE $F_t$  
\bigskip

1. An action of the Lie group SO(3,1) from the Virasoro algebra,

\bigskip

2. A lattice of inequivalent vacua,

\bigskip

3. A representation of the Kac-Moody Lie Algebra $ \tilde{E}_6$ by vertex
operators which shift vacua,

\bigskip

4. A family of fermionic vertex operators \cite{2reps} \cite{Fer}, 
which form a Clifford
algebra, because of the Fermi-Bose correspondence. These correspond to
representations of the double cover of the circle, so they have labels
from the generators of the $E_6$ lattice, or equivalently of
irreducible representations of A(4).}

\bigskip

In our search for the unified field theory, this appears as a sort 
of mathematical deus ex
machina. Coupling the EPRL model of quantum gravity to excitations
corresponding to representations of A(4) produces a model with
composite excitations corresponding to the gauge bosons of an $E_6$
grand unified model, together with fermions, which as we shall discuss
below closely resemble the fermions of the standard model.

We refer to the excitations of the field corresponding to the
irreducible representations of the group A(4) as {\bf tetrons}.

\bigskip

{\large \bf CONJECTURE:} {\it The EPRL model coupled to the tetron
  field gives a unified model which breaks to give the standard
  model}.

\bigskip

We detail the construction of the coupled model below.

A large body of recent work on grand unified models \cite{Gut} 
inspired by the 
neutrino
oscillations has focussed on symmetry groups which are a product of a
finite discrete group with a Lie group; and $A(4)\times E_6$ seems to
be the most successful of the possibilities. The natural coupling to
state sum models for quantum gravity and the construction of the $E_6$
action from vertex operators are the original parts of the
current proposal.

Discrete symmetry as a fundamental symmetry of nature is quite a
puzzle. In a continuum theory it would be hard to avoid domain
walls. The conceptual foundation of the BC and EPRL models, in which
spacetime is modelled by a simplicial complex and {\bf NOT } by a
manifold, is necessary to include discrete symmetry as in
this paper.

In the rest of this paper we give an introduction to the mathematical
structures which come together in our model, and describe how to
construct the coupled model. We hope it will provide the intuitive ideas 
and a guide to
the relevant literature. A more detailed study of the construction
will follow.

This paper is the result of connections appearing between different
fields not generally studied together, so the background material
really fits together in a web.

There is not a preferred order for reading the expository
sections of this paper. The reader is encouraged to read them as
needed. The two sections on the coupled model and the  boson-fermion
correspondence use all the
material. A physicist familiar with the dual model might want to go to
them first.

\bigskip

{\bf 1. THE BC AND EPRL MODELS}

\bigskip
The state sum models for quantum gravity \cite{BC} \cite{EPR}
are constructed on four
dimensional simplicial complexes, rather than smooth
manifolds. Classically the
geometry is specified by giving the bivectors (oriented area elements)
on the 2-faces of the complex. These can be quantized 
by representing them as balanced representations of SO(3,1). The
angles between the faces are described by the tensor product of the
representations on the faces. The structure of the models is based on
a state sum formula, in which the representations on the tetrahedra
bounding each 4-simplex are combined into a closed diagram or spin
network called a 15J symbol.

The EPRL model differs from the BC model in that the constraints are
imposed classically by requiring each tetrahedron to live in a space
slice of 
Minkowski space. Quantum mechanically, this is expressed by labelling
the faces of each tetrahedron by representations of SO(3), with an
intertwiner in the interior. The representations are then embedded in
representations of SO(3,1), the choice of which depends on the Immirzi
parameter $\tau$, and the representations of each tetrahedron are then 
acted on by a group element of SO(3,1) which is then integrated
over. We think of each tetrahedron as having a frame, and the
integrals over the group elements as a discrete version of the
integral over histories.

The tensor product on the representations and the action of the
group SO(3,1) are the structural elements of the model, to which any
matter fields must couple.  
\bigskip

{\bf 2. THE MATHEMATICAL STRUCTURE OF SPACETIME}

\bigskip
The state sum models for quantum gravity should not be thought of as
approximations corresponding to triangulations of an underlying smooth
manifold \cite{Me}. In quantum Physics coupled to relativity, such a point of
view is unphysical, because no information can be communicated from a
sub-Planck scale region. The attempt to recover a continuum limit is
the one unsuccessful part of the interpretation of the model, and is
naive.

Although the replacement of manifolds by simplicial sets is a
mathematically well understood point of view for geometry and
topology \cite{Book} \cite{Sul}, it seems very foreign to physicists.

Let us take some simple region of spacetime and compare two
descriptions of it, one as a subset of a smooth manifold, the other as a
simplicial complex, thought of as a discrete combinatorial object, not
a point set. Both of these have associated to them a differential
graded complex, the differential forms for the manifold, and the
cocycles for the complex. 

If we compute the cohomology of the region using both complexes, we
find there is an important difference: the differential forms can only
give vector spaces, while the cocycles can be calculated using integer
coefficients, and give finite group components to the cohomology as
well. This is known as the torsion. Thus finite group invariants
appear more naturally in the complex description.

Similarly, it is very easy to construct topological quantum field
theories from the representations of finite groups on a simplicial
complex, while a continuum Lagrangian is hard to imagine.

We interpret the discovery of discrete symmetry as a fundamental
symmetry in 
neutrino Physics as an indication that the simplicial complex
description is
favored by nature.
 
\bigskip

{\bf 3. NEUTRINO PHYSICS}

\bigskip
Durring the last few years, it has been established that neutrinos
have masses, and transform into one another \cite{exp}. The best explanation for
their masses is the seesaw model \cite{seesaw}. In this picture, neutrinos combine
Maiorana and Weyl masses. This explains the extreme lightness of the
left handed neutrinos in relation to the extremely heavy right handed
sterile neutrinos.  

This has the immediate consequence that any grand unified theory must
explain 
multiplets of 16 rather than 15 Weyl fermions in each generation. This
tends to favor SO(10) or $E_6$ models, rather than SU(5), in which the
15 form 10+5 as representations.

The 16 fermions of each generation have the natural structure of a
Grassman or Clifford algebra \cite{JB}; physicists sometimes describe it
equivalently as a fermionic state space. To see this, note that the
leptons of one generation form a left handed pair plus two right
handed singlets:

\bigskip

$ {\bf (e, \nu)_L +(e)_R +(\nu)_R} $.

\bigskip
These can be thought of as $ \Lambda (C^2)$, and in fact, it makes the
hypercharges come out right, because each basis vector of the $C^2$
has hypercharge 1/2, so the $\Lambda^0$ has hypercharge 0, the two
vectors of the $ \Lambda^1$ get hypercharge 1/2 and the vector in $
\Lambda^2$ gets hypercharge 1, explaining the charge of $e_R$. 

The quark sector has the same form except it is tensored with the
$C^3$ of color. Combining with the antiparticles gives us:

\bigskip

$ \Lambda (C^2) \otimes (C \oplus C^3 \oplus C^3 \oplus C )$.

\bigskip
which is isomorphic to $ \Lambda (C^5)$.

This form arises naturally in SO(10) GUT models, in which the
particles form a spin representation.
 
In order to find a unified theory which explains the fermions of the
standard model, we can equally well look for a fermionic state space,
a Grassman algebra, or a Clifford algebra, since the multiplication in
either algebra has no importance for us. The SO(10) unified theory
uses a spin representation or Clifford algebra; the SU(5) GUT uses a
Grassman algebra. In both cases the basis for the algebra can be taken
as the diagonal Cartan subalgebra of the Lie algebra.

\bigskip

{\bf 4. NEUTRINO OSCILLATIONS AND THE TRIBIMAXIMAL MATRIX}

\bigskip
The study of the neutrino oscillations revealed a number of
surprises. In particular, the mixing matrix was very different from
the one for quarks. The mixing matrix for the neutrinos is well
approximated by :
\bigskip

 \[ \left( \begin{array} {clcr}

\surd(2/3) & 1/ \surd 3 & 0\\

-1/ \surd 6 & 1/ \surd 3 & -1/ \surd 2 \\

-1/ \surd 6 &  1/ \surd 3 & 1/ \surd 2 \\

\end{array} \right) \]

\bigskip

which is called the tribimaximal matrix \cite{TBM}.

A factorization for this matrix was discovered, which led to an
explanation of it; namely that the leptons and Higgs particles had to
be assigned representations of the discrete group A(4) \cite{Ma}, while the
interaction between them had to be  invariant under the group.

The choice of A(4) was motivated by it having irreducible
representations 1, 1', 1'' and 3. There was an extensive process of
trial and error where finite groups of order up to 32 were tried and
mainly rejected. The other possibility is S(4), which is the symmetry
group of the cube.

Other than poetically \cite{PF}, no use was made of the fact that these groups
embed in SO(3) in the neutrino Physics literature.

The possibility that the discrete symmetry was a residual symmetry of
a Lie group was considered and rejected \cite{Rus}. There are just no field
configurations of fields in accessible representations which have
non-abelian residual symmetries.

So the discovery of the neutrino physicists is quite a puzzle for
theory.  It cannot be explained from any Yang-Mills type GUT.

\bigskip

\bigskip

{\bf 5. THE NEW GRAND UNIFIED MODELS AND THE PRINCIPLE OF UNIFICATION} 

\bigskip

Now the existence of sterile neutrinos with a mass in the range of
grand unification, together with the discrete symmetry proposal 
has led to a rather extensive study of GUT models
with a combination of discrete and continuous symmetry. In general,
they are rather successful phenomenologically. 

One author has published a metaanalysis of over 100 models of this
type \cite{exp}. The diversity of the subject is explained by the 
large number of
breaking schemes which are possible. According to his analysis, the
new models with $A(4) \times E_6$ symmetry best fit the empirical data.

However these models pose a philosophical problem:  the principle of
unification, as stated for example by Witten \cite{Wit} \cite{Wit2}. In the Yang-Mills
models, the principle was interpreted as meaning that the GUT group
had to be simple.  There did not turn out to be a four dimensional GUT
which unified the three generations of the standard model. This
motivated the turn to Kaluza-Klein models.

In the new situation, unification of the three generations comes from
the discrete ``horizontal'' symmetry. But how are we to unify a
discrete with a continuous group? 

In the case of A(4) and $E_6$, a very special mathematical connection
exists.

\bigskip

{\bf 6. THE MCKAY CORRESPONDENCE}

\bigskip
Mckay \cite{Mckay} discovered a strange correspondence between the discrete
subgroups of SO(3) and the simply laced simple Lie algebras, i.e. ADE
in the Cartan classification. 

The correspondence directly connects the representations of the
corresponding groups, or more precisely the ``binary''double cover
inside SU(2) of the finite
rotation groups with the Kac-Moody algebras corresponding to the simply
laced Lie algebras. When we draw the diagram whose vertices are the
irreducible representations of the dihedral group and whose edges
represent the tensor action of the two dimensional representation of
SU(2), we reproduce the extended Dynkin diagram of the loop group of
the corresponding ADE lie algebra. This was first discovered be
enumeration, and was considered a mystery.

The correspondence goes as follows:

\bigskip

$C_n$..........................$A_n$

$D_n$..........................$D_n$

A(4).........................$E_6$

S(4).........................$E_7$

A(5).........................$E_8$ 

\bigskip

where the cyclic groups correspond to the special linear groups, the
dihedral groups to the rotation groups, and the symmetry groups of the
platonic solids, tetrahedron, octahedron and icosahedron, correspond
to the exceptional Lie algebras.

Now the groups which appear in particle Physics: SU(2), SU(3), SU(5),
SO(10) and $E_6$ are all on our list.

Can we understand this correspondence in a way which helps us couple
excitations to quantum gravity, or discrete and continuous symmetries
to each other?

\newpage

{\bf 7. MODULE CATEGORIES}

\bigskip
In the categorification principle of the current author and Frenkel \cite{CF},
it was observed that structures on categories were formally similar to
algebraic structures, but with functors substituting for
functions. The category of representations of a finite group or 
Lie group is an important example; it is a ring category. The
operations $ \oplus$ and $ \otimes $ satisfy the axioms of a ring.

It is therefore natural to ask to classify the module categories over
the category of representations of a Lie group, SU(2) for example.

The answer turns out to be elegant and simple: the representation
categories of the subgroups of the Lie group are its module
categories, and there are no others. (The result for quantum groups is
almost the same) \cite{Eting}.

The category of representations of a subgroup is a module category
because the representations of the larger group can be considered as
representations of the subgroup, where they act by the ordinary tensor
product. 

So the list of finite subgroups above is also the list of module
categories over REP(SU(2)).

A module category can be coupled into a state sum model using the
module action analogously to the use of the tensor product in the
state sum 
model itself. We illustrate this below. So the small list of groups
above have a special affinity to the state sum models for gravity.

Can we act on them with the group SO(3,1) as in the EPRL model?
Can we obtain an action of the corresponding ADE group, $E_6$ in the
case of A(4)? In other words, can we construct the Mckay
correspondence in a physically interesting way to solve the problems of
the principle of unification and of coupling to the EPRL model?

\bigskip

{\bf 8. THE QUANTUM MCKAY CORRESPONDENCE OF FRENKEL JING AND WANG}

\bigskip

Frenkel Jing and Wang \cite{FJW} developed an approach to the Mckay
correspondence. In this section, we shall summarize it, and translate
it into physical terms.

The FJW construction begins with the finite group, then forms the
wreath products \cite{WB} of all degrees of it, and sums their
representation 
categories.

\bigskip

$ F_t = \bigoplus_n Rep[W( S_n, A(4))] $,

\bigskip

in the A(4) case, the other groups are treated analogously.

Here we are identifying a category with the space of formal sums of
its objects, i.e. its Grothendieck ring. This is an abuse of notation.

We need both to explain the wreath product and to show its connection
to bosonic quantisation, thus explaining why $F_t$ is regarded as a Fock
space.

The wreath product of $S_n$ with any finite group g is the semidirect
product of n copies of g with the permutation group $S_n$, where the
permutation group acts by permuting the copies of g. 

If we take n identical copies of a quantum mechanical system with
symmetry group g and impose permutation symmetry as well (symmetrize
the wave function), the total symmetry of the system will be the
wreath product.

Thus n-particle states of the symmetric system will be described by
representations of the n-fold wreath product.

We therefore regard $F_t$ as described above is a Fock space for
objects 
with A(4)
symmetry. We refer to it as tetronic Fock space.

Now $F_t$ has a very interesting structure, because the representation
categories of the wreath products are easier to describe all together
rather than separately. The representations of the n-fold wreath
product of a group correspond to the monomials of order n in the
irreducible representations of g with coefficients in C[z], so 
$F_g$ is a free polynomial
algebra with generators the irreducible representations of g.

This means that $F_t$ can be identified with the symmetric space on
the space of linear combinations of the irreducible representations of
$ \tilde{A}(4)$, the double cover of A(4) 

\bigskip

$F_t = Sym(Rep \tilde{A}(4) \otimes C[z]) $                (1)

\bigskip

Now this space can be identified with the Fock space of the dual model
with target space the torus $T_g$ obtained from quotienting the vector space
R(Rep g) by the integral lattice Rep(g) \cite{FK}. The powers of z in this
expression are the Fourier modes of the string, and correspond  in (1)
to the representations of the nth wreath product obtained by
symmetrizing n copies of a single irreducible representation.

Thus, the structure of a quantum string appears in $F_t$. The action
of the Virasoro algebra on the Fock space is available to us \cite{GS}. In
particular, the three generators $ L_{-1}, L_0, L_1, $ generate a copy
of the group SL(2,C) acting on $F_t$.  This gives us a natural way to
allow the SO(3,1) group operators in the EPRL model to act. Since we
can identify $F_t$ with a space of analytic functions on $CP^1$, the
standard formulas of Gelfand \cite{Gel} allow us to modify this to be 
a unitary
representation. The choice of the unitary representation will depend
on the value of the Immirzi parameter $\tau $.

The Fock space of the dual model has a family of vacua
corresponding to points in the lattice generated by the
representations of A(4) with the natural inner product of FJW.

We can think of these vacua as corresponding to homotopy classes of
embeddings of the string in the torus $T_t = R(Rep A(4))/Z(Rep(A(4))$. 

The vertex operators of this dual model give a representation of
the Kac-Moody algebra $ \tilde{E}_6$.

So we see the FJW construction gives us a unification of A(4) with
$E_6$, on which SO(3,1) has a natural unitary action.

\bigskip

{\bf 9. THE COUPLED MODEL}

\bigskip
Now we can see formally how we might construct a model which couples the EPRL
model to the tetron field. To each tetrahedron  in our discrete
spacetime, we associate a copy of $F_t$. We can think of this as a
tube passing through the tetrahedron joining the midpoints of the two
4-simplices it bounds. We act on the vector in $F_t$ by the element of
SO(3,1) associated to each side of the tetrahedron by the construction
of the EPRL model, using the action of SL(2,C) from the Virasoro
action on the string. We then trace together the five $F_t$ states
meeting on the center of each 4-simplex, using the operator assigned
to a five holed sphere by the conformal field theory associated to
$F_t$.

Picture this as tubes for each tetrahedron, joined together at the
center of each 4-simplex.

Let the representation of SO(3) on the intertwiner of each
tetrahedron in the EPRL model act quadratically on the vector in $F_t$
assigned to each tetrahedron. (Space tells matter how to move). We let
the intertwiner act quadratically because it is equivalent to letting
the four representations act because of the tensor law of a module
action, and is therefore independent of how the tetrahedron is
subdivided. 

Convergence issues for this formulation need to be studied.

We have constructed a sort of discretized string field theory. The
world sheets do not have a metric to integrate over ala Polyakov,
rather the string states propagate in the quantum geometry.

The bosons in the Lie algebra of $E_6$ appear as the constant
terms in the Kac-Moody action, given by vertex operators  as in
FJW. Aside from the problem of symmetry breaking, this gives us a
construction of the bosonic particles seen in nature.

Since the particles in nature appear in this theory as vertex
operators, rather than elements of the coupled state sum, the
development of this model into a computational technique is going to
require some difficult analysis.

\bigskip

{\bf 10. BOSON FERMION CORRESPONDENCE AND THE STANDARD MODEL}

\bigskip
The dual model associated to $F_t$, like many two dimensional quantum
field theories, has fermionic as well as bosonic excitations \cite{2reps}. The
fermionic excitations can be constructed using the isomorphism between
an infinite symmetric algebra and an infinite Clifford algebra. The
fermionic representation forms an infinite dimensional spinor.

The particle content of this representation is given by the constant
terms in the z expansion; it is just the Clifford algebra associated
to the Cartan subalgebra of $E_6$. This agrees with the form of the
fermions of the standard model discussed above, except that the rank
of $E_6$ is 6 instead of 5. This leaves us with a problem of reducing
the fermions by a factor of 2, which is part of the problem of working
out the symmetry breaking scheme.

The fermions do not appear as a complete 27 of $E_6$ as in
conventional GUTs. This is something of a head start towards symmetry
breaking. We give some preliminary thoughts on the problem of symmetry
breaking below.

If we construct the fermions as vertex operators \cite{Rec}, the 
vacua correspond to
embeddings of the double cover of the circle in $T_t$. Thus the vertex
operators would be labelled with representations of A(4) as well as
elements of the Clifford algebra. (The loops can go around a generator
of $T_t$ an odd number of times). This could explain why we seem to
see discrete symmetry on the fermions but not the bosons.

\bigskip

{\bf 11. SOME THOUGHTS ON SYMMETRY BREAKING}

\bigskip

The model we have constructed gives a plausible grand unified
theory. To test it, we need to specify a breaking mechanism which
reproduces the standard model. This part of the program is not done
yet.  Let us make some preliminary remarks.

The standard breaking mechanism in GUT theories is the Higgs
mechanism. As I write this, the data from CERN are ambiguous as to
whether the Higgs particle really exists.

Most scenarios starting from an $E_6$ GUT require several stages of
breaking. We can consider the possibility that some of the symmetry
breaking in our model is due to discrete processes in the state sum on
a simplicial complex that cannot be understood in a continuum
model. On the other hand, perhaps lower energy symmetry breaking, at
the weak scale for example, could be on a longer length scale for
which the continuum approximation is valid.

In this context, we make a rather strange observation. If we think of
the group A(4) as acting on the 2-sphere by rotations, the types of orbits
correspond to the classes of points on the tetrahedron. Now the
residual symmetries of the orbits are as follows: a vertex, or the
center of a face: $Z_3$, the midpoint of an edge: $Z_2$, a generic
point, the trivial group.

It is interesting to note that in the use of A(4) symmetry to explain
the tribimaximal mixing matrix, the states of the leptons and quarks
are not invariant under the full A(4) symmetry, but have residual
symmetries os $Z_3$ and $Z_2$.  

If we consider the action of the FJW constructions on these groups, we
obtain the Kac-Moody algebras over U(1), SU(2) and SU(3). In short the
residual symmetries of $E_6$ which must appear in the standard model.

Could some phase transition in the coupled model produce an alignment
of the quantum geometries of the tetrahedra,
in which only the residual symmetries appear at low energy?

\bigskip

{\bf 12. PHYSICAL CONSEQUENCES OF THE MODEL}

\bigskip

The proposal that physical particles correspond to different vacua of
an underlying quantum field of tetrons is a major departure. Could it
have observable consequences?

One obvious possibility is that the tetron field is responsible for
the dark matter and energy. The fact that we cannot directly observe
it makes the testing of this possibility difficult, but perhaps the
interaction between the vertex operators and basic generators of $F_t$
will tell us something.

A very challenging problem facing any attempt at a unified
theory is called the hierarchy problem. It is hard to
explain why the coupling scale at which mass is generated is so many
orders of magnitude below the natural scales of the model, the Planck
or unification scales.

While the coupled model proposed here is in much too early a state of
computational analysis to attack this problem directly, a natural
hypothesis suggests itself:

\bigskip

{\bf CONJECTURE} {\it Since the physical fermions and bosons present
  themselves as vacuum changing operators on the tetron field, the
  couplings between them are Vanderwaals type interactions of higher
  order, and are therefore extremely weak.}
\bigskip

Given the difficulty of the hierarchy problem, this
suggestion is at least worth investigation.
 
\bigskip

{\bf 13. REFLECTIONS}

\bigskip

To summarize: the discrete symmetry discovered in neutrino Physics is
related to the Lie Group $E_6$ by the Mckay correspondence. The FJW
version of the Mckay correspondence allows us to construct the Lie
group as vertex operator algebras on a quantum field of
representations of the discrete group. The symmetry breaking of the
discrete group corresponds by the Mckay correspondence to the pieces
into which $E_6$ must break to give the standard model. The fermionic
excitations of the FJW model form a Clifford algebra, as do the
fermions of the standard model with the new neutrino Physics included.

The discrete group which appeared in neutrino Physics is almost unique
in coupling to the tensor structure in the new state sum models for
gravity.

These connections were hard to make because they are in different
fields not generally studied together. Once assembled, they seem to
fit together naturally.

The tetron fields can really be thought of as a kind of quantum
geometry. They can be constructed in terms of the Hilbert schemes
\cite{Nak} \cite{Nak2}, 
which are just the spaces of n-tuplets of points, in an orbifold
formed by quotienting the 3-sphere by the action of the group
A(4). Thus the coupled model is quite similar to the construction of
the BC or EPRL model by using representations as quantizations of
geometrical objects.

If a natural breaking scheme can be found to give the standard model
from the coupled model, then calculations of quantities such as masses
from particle Physics could  provide confirmation of the
coupled model, and indirectly of the EPRL model for quantum gravity as well.

It was quite a surprise to have the string appear in this theory,
which started from a completely different program. Rather than ending
up with an almost infinite landscape as in Kaluza-Klein theories, we
get an essentially unique theory, which relates fairly directly to the
standard model.

Embedding the string field in a discrete model for spacetime removes
the difficulties that beset string field theories in a continuum.

There is no longer the integration over worldsheet metrics which leads
to bad behavior on moduli space in the Polyakov string; rather the
dual models couple to the quantum geometry of the EPRL model itself.

So perhaps the intuition at the bottom of string theory finds a
suitable setting in our coupled model. Indeed, physical particles
correspond to vertex operators in our picture, while the 10
dimensional theories are more subtle.

\bigskip

{\bf 14. ONGOING PROGRAM}

\bigskip

In order to turn this concept into a working physical theory, a
computational program will be necessary. The first phase will be to
find an approximation scheme for the coupled model. It seems plausible
that the asymptotic analysis of the EPRL model \cite{B} could be
coupled to the theory of vertex operators to do this.

The objective of this phase would be to reproduce quantum
probabilities in an $E_6$ GUT model. The Yukawa couplings could be
approximated in the background of a quantum 4-simplex, for example.

The second stage would require us to identify a breaking scheme, and
would ultimately be expected to reproduce the masses of the standard
model.

In short, the program is challenging but not hopeless.

\bigskip
{\bf ACKNOWLEDGEMENTS:} The ideas in this paper were developed in a
seminar in KSU. Dany Majard and David Yetter contributed many useful
questions. Part of the work was done while the author was visiting at
Paris VII. The author wants to thank Marc Lachieze-Rey for his
hospitality and many helpful remarks. Martin Bucher made many helpful
inputs durring some of the research.

\end{document}